\documentclass[twocolumn,aps,prl,showpacs,floatfix]{revtex4}
\usepackage{graphicx}
\usepackage{amsmath}
\usepackage{hyperref}
\usepackage{latexsym}
\usepackage{color}
\usepackage{epsfig}


\begin{document}

\title{Localization of bosonic atoms by fermionic impurities  in a 3d optical lattice}

\author{S. Ospelkaus, C. Ospelkaus, O. Wille, M. Succo, P. Ernst, K. Sengstock and K. Bongs}
\affiliation{Institut f\"ur Laserphysik, Luruper Chaussee 149, 22761~Hamburg / Germany}

\begin{abstract}
We observe a localized phase of ultracold bosonic quantum gases in a 3-dimensional optical lattice induced by a small contribution of fermionic atoms acting as impurities in a Fermi-Bose quantum gas mixture. In particular we study the dependence of this transition on the fermionic $^{40}$K impurity concentration by a comparison to the corresponding superfluid to Mott insulator transition in a pure bosonic $^{87}$Rb gas and find a significant shift in the transition parameter. The observed shift is larger than expected based on a mean-field argument, which is a strong indication that disorder-related effects play a significant role.
\end{abstract}


\maketitle

Ultracold atomic gases have already shown to be versatile model systems for quantum phenomena in many areas of physics. Prominent examples are the recent demonstration of the superfluid to Mott insulator transition \cite{1} and fermionic superfluidity in the BEC-BCS crossover regime \cite{2,3,4,5,6}. These phenomena of so-called ``strongly correlated physics'' originate from the rich field of transport and localization in condensed matter systems with high relevance for technical applications. Realizations of such systems with ultracold atoms in optical lattices are a breakthrough as they offer a ``defect-free'' system as well as in situ control over tunneling and interaction parameters, allowing a direct quantitative comparison to theory. However, in contrast to single-component quantum gases, real condensed matter systems are not defect-free and impurities play a crucial role, namely causing important localized phases connected to charge-density waves, disorder or percolation. Thus, the realization of impurity physics with quantum gases is of high relevance. Fermi-Bose mixtures in 3-dimensional optical lattices provide a versatile model system for many disorder- and impurity-related effects.

Interest in this novel strongly correlated mixed statistics system has been driven by investigations of new Fermi-Bose correlations and their consequences \cite{14,15,16,17,18,19,20}. In the extreme case of pairing of fermions with one or more bosons a whole zoo of new quantum phases of these ``composite fermions'' has been predicted \cite{11}, and connections to superconductivity \cite{14,15} but also ``exotic'' systems like neutron stars and quark-gluon plasmas \cite{13} suggest fundamental relevance for a broad range of phenomena. Even before such atom pairs form, Fermi-Bose correlations are predicted to become manifest in the formation of charge density waves \cite{7,8,9} or supersolids \cite{17}, but also in polaron-related physics of fermions dressed with bosons \cite{7} or percolation \cite{10}. Such disorder-driven localization phenomena are present and intensely discussed in many systems, from Anderson localization in condensed matter and light localization in random media to seismic wave dynamics and even disturbances in traffic flow.

\begin{figure}[tbp]
\begin{centering}
\leavevmode
\resizebox*{1\columnwidth}{!}{\includegraphics{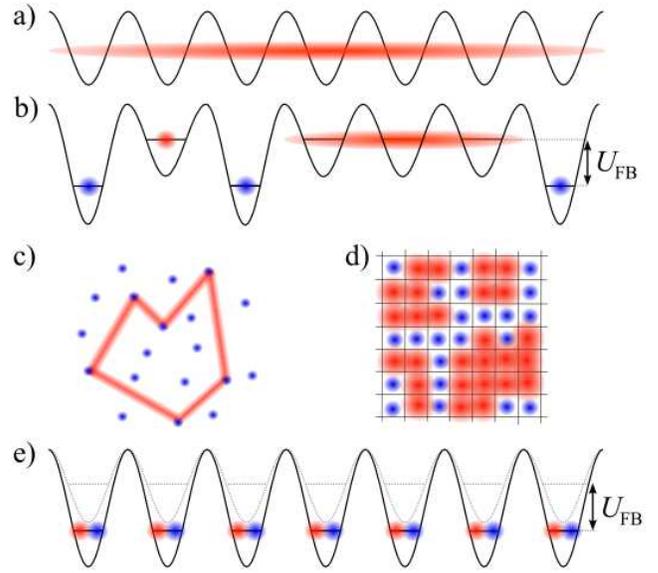}}
\end{centering}
\caption{Schematic localization scenarios. 
{\bf a}. Pure bosonic superfluid in an optical lattice. 
{\bf b}. Shift of the effective potential depth due to fermionic impurities. 
{\bf c}. Localization by interfering paths of the bosonic wavefunction scattered by randomly distributed fermionic impurities. 
{\bf d}. Localization due to percolation. A random fermion distribution hampers the establishment of a coherent connection and causes the localization of bosonic ensembles in superfluid ``islands''. 
{\bf e}. Mott insulator transition induced by a uniform distribution of attractive fermionic impurities, resulting in an effectively deeper lattice potential for the bosons.} \label{fig1}
\end{figure}

In this letter, we present first studies on Fermi-Bose mixtures confined in 3-dimensional optical lattices where fermionic atoms act as impurities. We observe the onset of a localized phase in the bosonic $^{87}$Rb component induced by the fermionic $^{40}$K impurity atoms. We find a fermion concentration dependent shift with respect to the quantum critical point of a pure bosonic ensemble which is incompatible with simple mean-field arguments. This indicates that the observed transition is driven by disorder-related many-body effects. The measurements presented in this letter are the first to combine three-dimensional crystals made of light with quantum particles of mixed statistics. On the one hand, this novel system is a realization of the Bose-Fermi Hubbard Hamiltonian \cite{Albus2003,18}; on the other hand, it offers further potential for controlled investigations of impurity-induced phases as well as studies on pairing between bosons and fermions leading to composite fermions \cite{11,12,13} and BCS-like boson-assisted fermion pairing \cite{14,15}

We first consider intuitive localization scenarios for Fermi-Bose mixtures confined in 3-dimensional optical lattices, starting with a pure bosonic superfluid (Fig.~\ref{fig1}a). Adding fermionic impurities and considering the attractive $^{40}$K --- $^{87}$Rb Fermi-Bose interaction energy as an additional potential for the bosons, the ``defects'' caused by the fermionic impurities can be described by a local change of the effective optical lattice depth for the bosons due to the interparticle interaction (Fig.~\ref{fig1}b). If the energy level shift caused by the interaction energy is large enough, the superfluid bosonic wavefunction will not extend into this defect region, but will be scattered by the impurity. If scattering becomes frequent, interference effects along a closed scattering path are predicted to suppress transport and lead to a localization scenario similar to Anderson localization \cite{21} (Fig.~\ref{fig1}c). A further increase in the impurity density will lead to the formation of ``forbidden walls''. Once the walls in this quantum percolation scenario lead to a sufficiently complicated labyrinth like structure for the bosonic wavefunction, a single coherent superfluid phase can no longer be sustained and several separated domains will be formed (Fig.~\ref{fig1}d). For the maximum filling of one fermion per lattice site the localized phase becomes comparable to a pure bosonic Mott-insulator but now the transition is shifted by the interaction energy with the fermionic impurity (Fig.~\ref{fig1}e). Access to these various regimes in the experiment can be controlled by varying both crystal depth and fermionic impurity concentration. 

Our experimental procedure for creating degenerate mixtures has been described previously \cite{25,jmo}. In brief, we create Fermi-Bose mixtures by first collecting and precooling bosonic $^{87}$Rb and fermionic $^{40}$K atoms using a combined 2d-3d MOT laser cooling system. After transfer to a Ioffe-Pritchard type magnetic trapping potential we perform forced RF evaporative cooling of the rubidium component in the $\left|2,2\right>$ state (which in turn sympathetically cools the potassium atoms in the $\left|9/2,9/2\right>$ state) until the mixture is close to quantum degeneracy. At this point we add a dipole trapping potential in order to increase the axial trapping frequency from 11 to $\approx$50 Hz (this and all subsequent trap frequencies correspond to the $^{87}$Rb component) and at the same time reduce the radial trapping frequencies from 250 Hz to 150 Hz, resulting in a significant decrease of the trap aspect ratio. Having created a quantum degenerate Fermi-Bose mixture by a last evaporation step in this combined trap, either this mixture is used for further experiments or the fermions are removed with a short resonant light pulse, leaving a pure bosonic sample as reference. The ultracold mixture of about $10^5$ bosonic $^{87}$Rb atoms with no discernible thermal cloud contains a variable 5-20\% fermionic $^{40}$K atom impurity component.

\begin{figure}[tbp]
\begin{centering}
\leavevmode
\resizebox*{1\columnwidth}{!}{\includegraphics{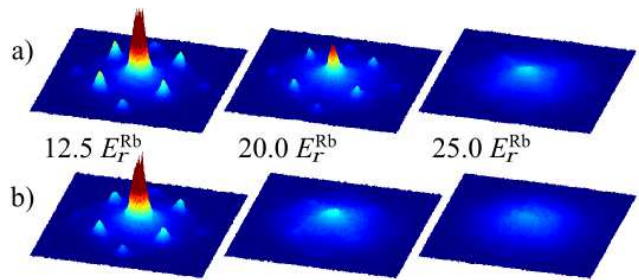}}
\end{centering}
\caption{Time of flight absorption images of the bosonic component 15 ms after
switching off the lattice and trap potentials. (The colors and pseudo-3d
representation encode the particle column density integrated along the imaging
direction). 
{\bf a}. Pure bosonic ensemble for three different lattice depths.
{\bf b}. Fermi-Bose mixture for the same lattice depths and 20\% fermionic impurity concentration.
The reduction of interference contrast visible in the images shows the onset of localization, which is shifted to lower lattice depth in the presence of fermionic impurities.} \label{fig2}
\end{figure}

The optical lattice potential is created by three orthogonal and retroreflected laser beams (waists 82/92 $\mu\textrm{m}$ in the radial directions and 55 $\mu\textrm{m}$ in the axial direction with respect to the original trap) with mutually orthogonal polarizations and a relative frequency difference of about 10MHz. The lattice is operated at a wavelength of 1030 nm, creating a simple cubic crystal with a lattice constant of 515 nm \cite{VersaDisk}. Due to the additional harmonic confinement by the magnetic trap and the Gaussian lattice laser profiles, the mixture occupies a few ten thousand lattice sites with an occupation rising from 0 in the outer regions to 1 fermion and $>$5 bosons per site at the center. Note that we have chosen parameters such that the fermionic impurities always stay within the bosonic cloud. After sudden switch-off of the lattice potential and a period of free expansion of 15-20 ms we record the interference pattern which builds up in the density distribution of the bosonic component \cite{FermiImage}. In order to map the localization transition we repeat this procedure for several different final lattice depths. 

Fig.~\ref{fig2} shows the evolution of the boson interference pattern for about 20\% impurity concentration (bottom row) in comparison to a pure bosonic sample created under the same experimental conditions but with a removal of the fermionic atoms just before the lattice ramp-up sequence (top row). The lattice depths are given in units of the recoil energy for the $^{87}$Rb component $E_r^{\mathrm{Rb}}=(h^2k^2)/(8\pi^2m_{\rm Rb})\approx h\cdot2.14\,\textrm{kHz}$ where $k$ is the lattice wavenumber. The loss of interference contrast marking the breakdown of long range order is clearly visible in both cases. In case of the pure bosonic gas, the loss of coherence is due to the well-known superfluid to Mott-insulator phase transition \cite{jaksch,1} which occurs as a result of competition between the minimization of kinetic energy, parametrized by the tunneling matrix element $J$ which tends to delocalize the atomic wavefunction over the crystal and the minimization of interaction energy $U$ (Fig.~\ref{fig2}a). As can be clearly seen from Fig.~\ref{fig2}b, the presence of fermionic impurities induces a loss of coherence at much lower lattice depths than for a pure BEC. 

\begin{figure}[tbp]
\begin{centering}
\leavevmode
\resizebox*{1\columnwidth}{!}{\includegraphics{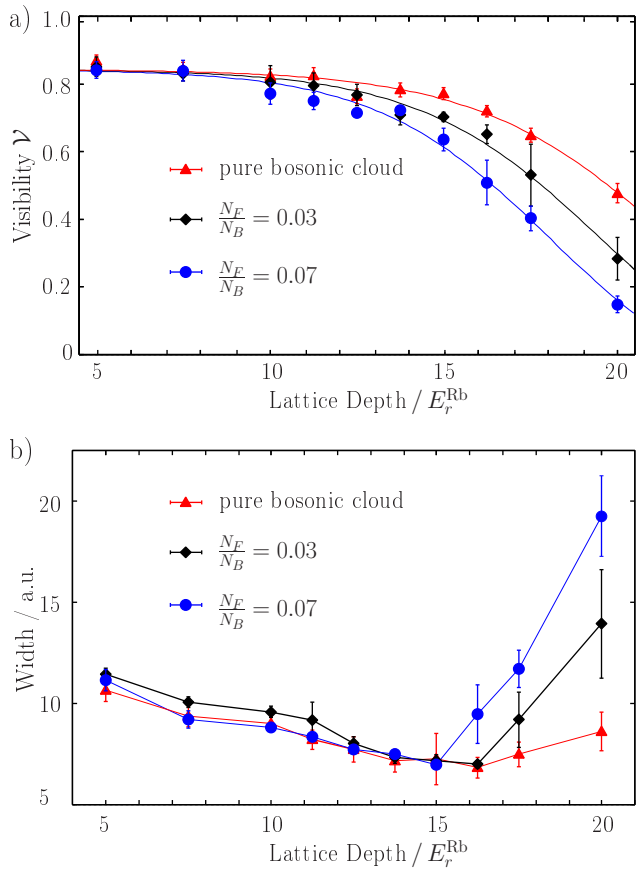}}
\end{centering}
\caption{Visibility $\mathcal{V}$ (a) and central peak width (b) of the bosonic interference pattern for different fermionic impurity concentrations.} \label{fig3}
\end{figure}

In the following, we will study this impurity-induced phenomenon quantitatively. Taking the interference contrast in the above pictures as a measure of coherence and putting it into mathematical terms, one can define the visibility of the matter wave interference pattern as \cite{22} $\mathcal{V}=(N_{peaks}-N_{int})/(N_{peaks}+N_{int}).$ Here $N_{peaks}$ denotes the sum of the number of atoms within the first order interference maxima, while $N_{int}$ is the sum of the number of atoms in equivalent areas at intermediate positions between the maxima. In addition, we extract complementary information from the width of the central interference peak as a measure of the correlation length of the bosonic system \cite{1,23}. Fig.~\ref{fig3} shows sample data comparing the behavior of ensembles with 3\% and 7\% impurity concentrations to a pure bosonic ensemble. The onset of localization marked by a loss of interference contrast and a decrease of correlation length is clearly visible in the visibility curve as well as in the width data. Fig.~\ref{fig3} clearly shows that the onset of the localized phase is shifted by an amount depending on the impurity concentration. Already for 3\% impurity concentration, the visibility data shows a significant shift. The corresponding width data, however, essentially exhibits a steeper slope in the localized phase with no pronounced shift as compared to the pure bosonic system. This indicates a qualitative difference in the type of transition for the pure bosonic system and the system with fermionic impurities.

In order to assess the size of the shift, one can calculate the value of the on-site Fermi-Bose interaction matrix element 
$$U_{\mathrm{FB}}=\frac{2\pi\hbar^2 a_{FB}}{\mu}\cdot\int w_F(x)^2 w_B(x)^2\,d^3x=1.1E^{\mathrm{Rb}}_r$$
The value on the right-hand side is calculated in the tight binding limit at the localization transition point for our lattice parameters, based on a scattering length of $a_{\mathrm{FB}}=-205(7)a_0$ \cite{24}. This matrix element characterizes the additional potential felt by bosonic atoms due to the presence of a fermionic impurity atom in a single potential well. It thus corresponds to the expected shift of the localization transition for unity fermionic filling in the simple mean field picture shown in Fig.~\ref{fig1}e. 

\begin{figure}[tbp]
\begin{centering}
\leavevmode
\resizebox*{1\columnwidth}{!}{\includegraphics{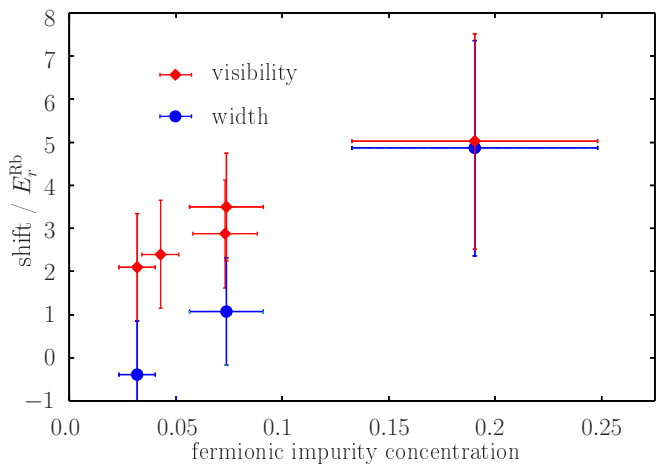}}
\end{centering}
\caption{Observed shift of the localization transition as a function of fermionic impurity concentration} \label{fig4}
\end{figure}

We use two different methods to extract the shift of the localization transition from our experimental data. For the visibility data, we use a phenomenological fit function
$$\mathcal{V}(s)=\frac{\mathcal{V}(0)}{1+\mathrm{Exp}(\alpha\cdot(s-s_{\mathrm{crit}}))}$$
where $\mathcal{V}(0)$ is the visibility for shallow lattices, $s$ is the lattice depth in units of $E_r^{\mathrm{Rb}}$, $s_{\mathrm{crit}}$ is a measure for the onset of the localization transition and $\alpha$ is an additional fit parameter. For data as in Fig.~\ref{fig3}, we compare $s_{\mathrm{crit}}$ for a pure bosonic sample recorded under the same experimental conditions to the impurity induced transition point and extract the shift of the transition. Resulting shifts are plotted as a function of impurity concentration in Fig.~\ref{fig4}. The corresponding width data is analyzed by extracting the intersection point of two linear fits to the descending and ascending branches of the data. As can be seen from Fig.~\ref{fig4}, an increasing impurity concentration leads to a considerable shift of the localization transition, significantly larger than the on-site Fermi-Bose interaction energy of $U_{\mathrm{FB}}$. For 20\% impurity concentration, the shift is on the order of 5$E_r^{\mathrm{Rb}}$, about 5 times $U_{\mathrm{FB}}$. Our data is therefore clearly incompatible with the above mean-field argument even for unity fermionic filling \cite{higherbands}, and indicates additional effects like disorder. At the same time, for smaller impurity concentrations, the shift extracted from the width is always smaller than the visibility data. Thus our data suggest that global phase coherence is reduced with fewer impurities than the correlation length, also indicating the presence of disorder effects. 

We have carefully checked temperature and ramp speed dependence as well as heating and loss effects and found that despite some losses in fermion number compatible with three body recombination \cite{25} and mild heating, the impurity effects caused by the fermions are clearly identifiable. Furthermore the localization transition is reversible, i.e. the visibility is nearly fully recovered, when the lattice depth is ramped back down, which shows that non-adiabatic effects play a minor role. 

Fermi-Bose mixtures in a three-dimensional optical crystal as presented in this letter open many new opportunities for the experimental investigation and understanding of localization phenomena on various grounds. Future studies, e.g. using second order correlations or Bragg spectroscopy, will concentrate on the further isolation and identification of the underlying phases and associated phase transitions in this rich system.

\begin{acknowledgments}
We acknowledge fruitful discussions with M. Cramer and G. Shlyapnikov as well as financial support by Deutsche Forschungsgemeinschaft (SPP 1116).
\end{acknowledgments}


\begin{thebibliography}{28}
\bibitem{VersaDisk}
The light for for the optical lattice is provided by a commercial Yb:YAG disc laser with a maximum output power of 20W which is frequency stabilized to a reference cavity with a residual relative linewidth of 20kHz, measured over 50ms.
\bibitem{FermiImage}
The fermionic component is simultaneously imaged after 3-5 ms time of flight and is used to determine the impurity concentration.
\bibitem{higherbands}
Population of higher bands by fermionic atoms could explain a larger shift. Within our experimental resolution, we have not observed any occupation of higher bands by fermions.
\end{thebibliography}
\end{document}